\documentstyle[aps]{revtex}

\def\dofig#1#2{\epsfysize=#1 \centerline{\epsfbox{#2}}}
\input{epsf}
\begin{document}
\title{Encryption with Delayed Dynamics}
\author{Toru Ohira}
\address{
Sony Computer Science Laboratory\\
3-14-13 Higashi-gotanda, Shinagawa,\\
 Tokyo 141, Japan\\
(Sony Computer Science Laboratory Technical Report: SCSL-TR-98-014)
}

\date{\today}
\maketitle
\begin{abstract}
We propose here a new model of encryption
of binary data taking advantage of the complexity
associated with delayed dynamics.
In this scheme, the encryption process 
is a coupling dynamics with various time delays between different
bits in the original data. It
is shown that decoding of the encrypted data is
extremely difficult without a complete 
knowledge of coupling manner with associated delays
for all bits of the data.
\end{abstract}
\vspace{1em}

Complex behaviors due to time delays
are found in many natural and artificial systems. 
Some examples are delays in bio-physiological
controls ({\it 1}, {\it 2}), and signal transmission delays 
in large--scale networked or distributed information
systems (See e.g. ({\it 3}, {\it 4})). 
Research on systems or models with delay has
also been carried out in the fields of 
mathematics({\it 5}, {\it 6}), artificial neural networks ({\it 7}, {\it 8}),
and in physics ({\it 9}, {\it 10}, {\it 11}).
This series of research has revealed that time delay can
introduce surprisingly complex behaviors to otherwise
simple systems, because of which delay
has been considered an obstacle from the
point of view of information processing.

In this paper, however, we actually take advantage of this
complexity with delayed dynamics and incorporate it
into a new model of encryption.
The encryption process is identified with 
a coupling dynamics with various time delays between different
bits in the original data. 
We show that the model produces a  complex behavior with
the characteristics needed for an encryption scheme.

Let us now describe the encryption model in more detail. 
${\bf{S}}(0)$ is the original data of $N$ binary bits, whose
$i$th element $s_i(0)$ takes values $+1$ or $-1$.
The delayed dynamics for the encryption can be specified
by a key which consists of the following three parts:  
(1) a permutation ${\bf P}$ generated from $(1,2,3,..,N)$, 
(2) a delay parameter vector $\vec{\tau}$ which consists of $N$ positive
integers, and 
(3) number of iterations of the dynamics $T$.
Given the key $K = ({\bf P}$, $\vec{\tau}$, $T)$, the dynamics
is defined as
\begin{equation}
s_i(t) = (-1)\times s_{p_i}(t-\tau_i), 
\end{equation}
where $p_i$ and $\tau_i$ are $i$th element of ${\bf P}$ and $\vec{\tau}$, 
respectively. (If $t-\tau_i < 0$, we set $t - \tau_i = 0$.)
In Figure 1, this dynamics is shown schematically.
The state of the $i$th element of ${\bf{S}}(t)$ is given
by flipping the state of the $p_i$th element of ${\bf{S}}(t-\tau_i)$.
Thus this dynamics causes interaction between $N$ bits of 
the data in both space and time. The encoded state 
${\bf{S}}(T)$ is obtained by applying this operation
of equation (1) iteratively $T$ times starting from
${\bf{S}}(0)$.
\begin{figure}[h]
\dofig{1.8in}{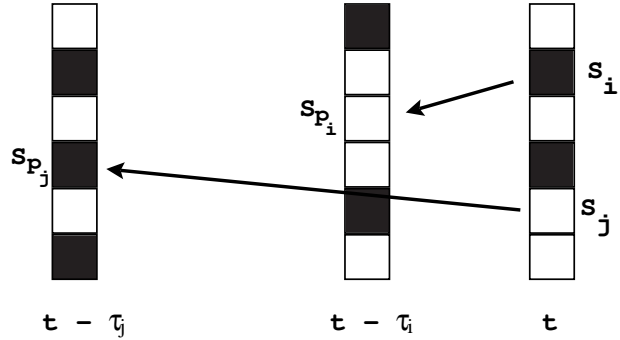}
\caption{
Schematic view of the model dynamics. 
The state of the $i$th element of ${\bf{S}}(t)$ is given
by flipping the state of the $p_i$th element of ${\bf{S}}(t-\tau_i)$.
}
\end{figure}

We investigate numerically the nature of the delayed dynamics 
from the perspective
of measuring the strength as an encryption scheme.
First, we examine how the state ${\bf{S}}(t)$ evolves
with time. In Figure 2 , we have shown an
example of encoded states with different $T$ using the
same ${\bf P}$ and $\vec{\tau}$ for a case of $N=81$.   
To be more quantitative,
we compute the following quantity as a measure of difference
between two encoded states at different times $t$ and $t_f$:
\begin{equation}
Y(t)={1 \over N} \, \sum_{i=0}^NS_i(t)S_i\left(t_f\right)
\end{equation}
A typical example is shown in Figure 3. We note 
that the dynamics of our model has a property of
occasionally very similar, but not exactly the same, states appearing
indicated by sharp peaks in the figure.
Except around these particular points, however,  we generally
obtain rather uncorrelated encoded states (i.e., $Y \approx 0$) with
different iteration times. This is a desirable property of the model as
an encryption scheme: the same state can be encoded into
uncorrelated states by changing $T$.
\begin{figure}[h]
\dofig{3.5in}{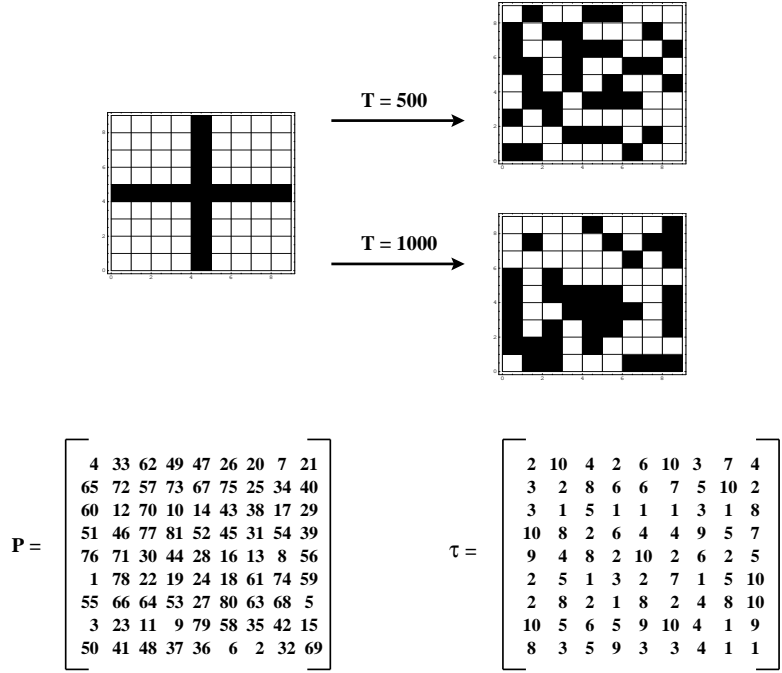}
\caption{
Examples of encoding with the model dynamics from an initial state to 
$T=500$ and $T=1000$ with ${\bf P}$ and $\vec{\tau}$
}
\end{figure}

\begin{figure}[h]
\dofig{2.0in}{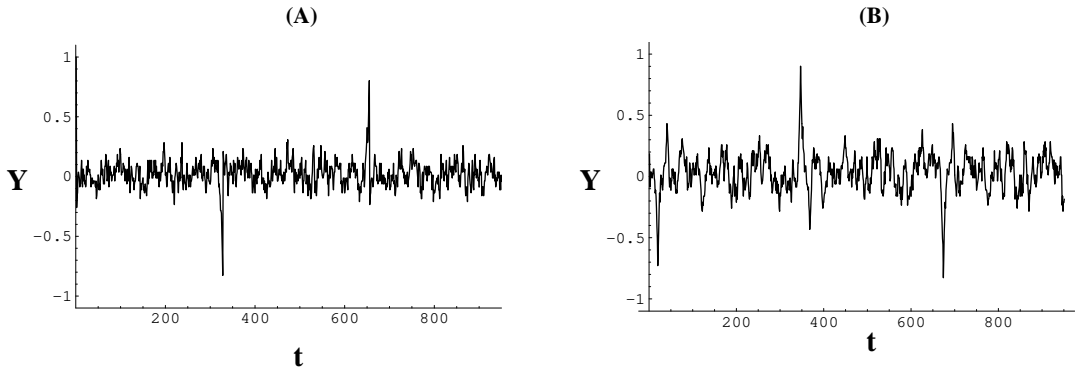}
\caption{
Examples of the correlation $Y$ between encoded states at different time steps evaluated by
equation (2).  Cases with (A) $t_f = 0$ and (B) $t_f = 1000$ are plotted
with the initial state and ${\bf P}$ and $\vec{\tau}$ the same as in Figure 2.
}
\end{figure}

Next, we investigated the effect of a minor change of ${\bf P}$ and $\vec{\tau}$
on the model dynamics. Starting with the same initial condition,
we evaluate how two states ${\bf{S}}(t)$ and ${\bf{S'}}(t)$ are encoded with slightly different 
${\bf P}$ and ${\bf P'}$, respectively, by computing
\begin{equation}
X(t)={1 \over N} \, \sum_{i=0}^NS'_i(t)S_i\left(t\right)
\end{equation}
A representative result is shown in Figure 4.  The same evaluation with
$\vec{\tau}$ and $\vec{\tau' }$ is shown in Figure 5.
These graphs indicate that if we take sufficiently large $T$, the
same state can evolve into rather uncorrelated states
with only a slight change of ${\bf P}$ and $\vec{\tau}$. 
This again is a favorable property in the light of encryption.
It makes iterative and gradual guessing of ${\bf P}$ and $\vec{\tau}$
in terms of their parts and elements very difficult: a nearly correct guess
of the values of ${\bf P}$ and $\vec{\tau}$ does not help in decoding.
\vspace{5em}

\begin{figure}[h]
\dofig{3.5in}{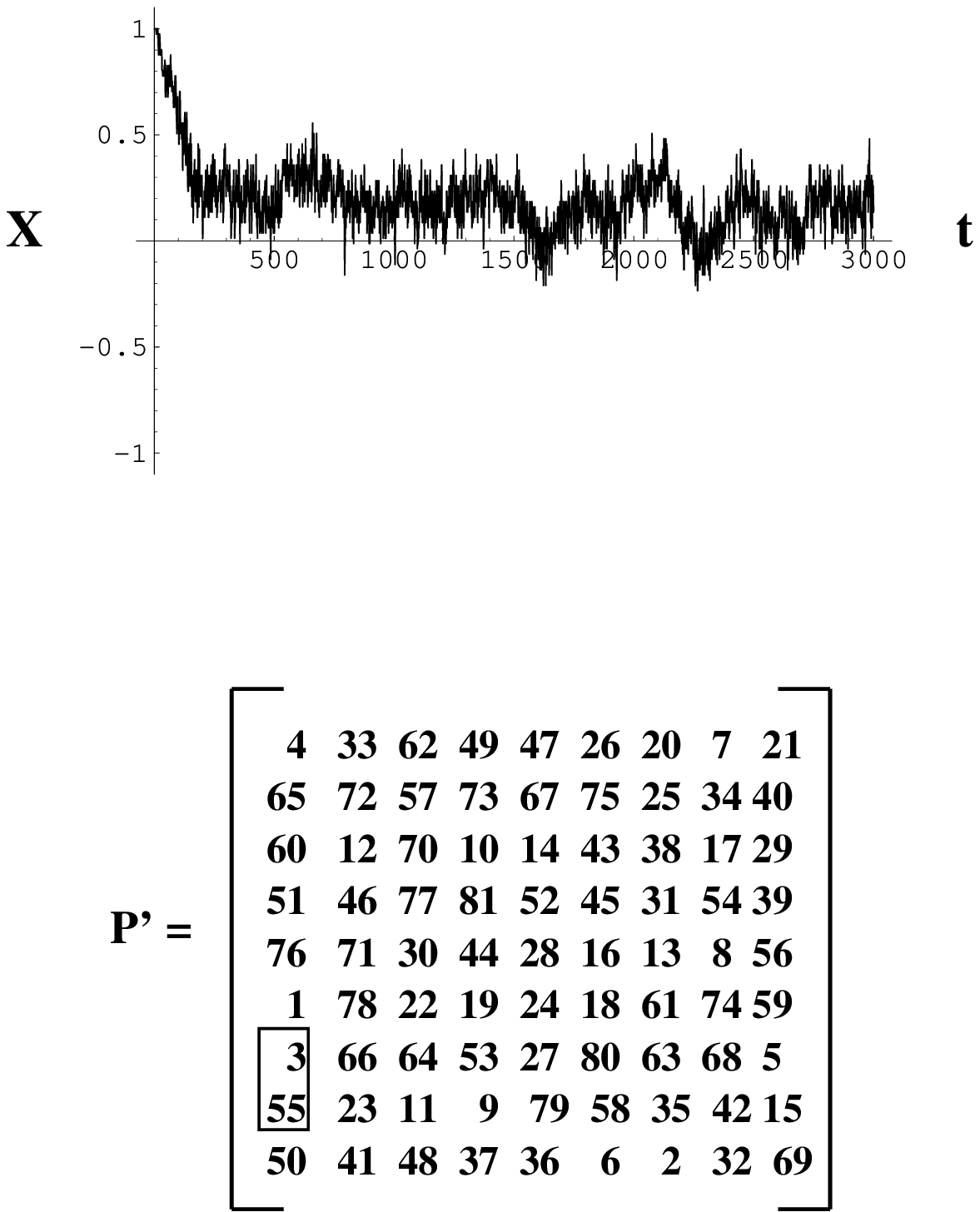}
\caption{
Example of the correlation $X$ evaluated by
equation (3) between two states encoded by slightly different
${\bf P}$ and ${\bf P'}$ .  The difference of ${\bf P'}$ from ${\bf P}$
is indicated by a box. 
The initial state and ${\bf P}$ and $\vec{\tau}$ are the same as in Figure 2.
}
\end{figure}

\begin{figure}[h]
\dofig{3.5in}{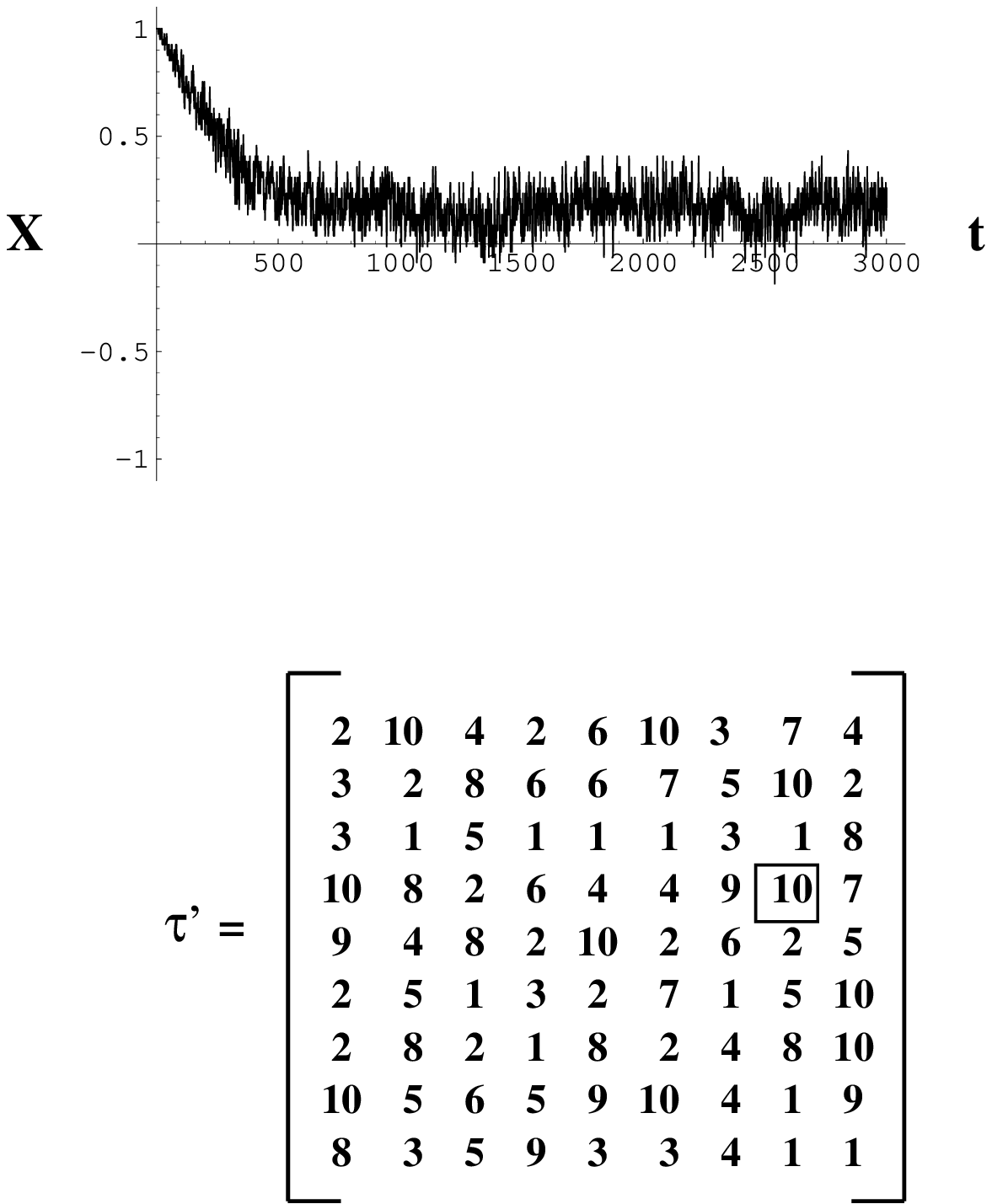}
\caption{
Example of the correlation $X$ evaluated by
equation (3) between two states encoded by slightly different
$\vec{\tau}$ and $\vec{\tau' }$.  The difference of $\vec{\tau' }$ from $\vec{\tau}$
is indicated by a box. 
The initial state and ${\bf P}$ and $\vec{\tau}$ are the same as in Figure 2.
}
\end{figure}

With these properties of the model, an exhaustive search appears to be
the only method for guessing the key.  Even if  one knows $N$ and 
$\tau_{max}$, the largest element in $\vec{\tau}$,  one is still required to
search for the correct key from among $(N!)(\tau_{max})^N$ combinations and to guess $T$. 
The commonly used DES (Data Encryption Standard) employs $2^{56}$ bit
keys ({\it 12}) . We can obtain a similar order of difficulty 
with rather small values of $N$ and $\tau_{max}$; 
for instance,  $N \approx 11$ and $\tau_{max} \approx10$ ({\it 13}). 

There are different methods possible for using this model for a secure communication
between two persons who share the key.  One example is that
the sender sends a series of encoded data in sequence for the interval
between $T$ and $T+\tau_{max}$ (or longer). 
The receiver can recover 
the original data from this set of encoded data by applying a reverse
dynamics with the key.   In a situation where the data sent is a choice out
of multiple data sets known to the receiver,  the receiver can
run the encryption dynamics to the entire sets with the key for
case matching.

As in many dynamical systems in the presence of delay, 
the behavior of the model
presented here is not analytically tractable as it stands.  
The model can be viewed as an extension of iterative function on a vector, 
or as an extension of cellular automata,
with different delays for each element.
Analytical and numerical investigation of the model from these perspectives
as well as implementation of the model into a software tool
is currently underway.
It is hoped that the model presented here can serve to call attention to  
the issue of encryption from the standpoint of delayed dynamics studies taking
place in various scientific disciplines.
\vspace{2em}

\begin{center}
{\bf REFERENCES AND NOTES}
\end{center}

\begin{itemize}
\begin{enumerate}

\item
M. C. Mackey and L. Glass, {\it Science} {\bf 197}, 287 (1977);

\item
A. Longtin and J. Milton, {\it Biol. Cybern.} {\bf 61}, 51 (1989); 

\item
P. Jalote, {\it Fault Tolerance in Distributed Systems.}  
(Prentice Hall, NJ, 1994).

\item
N. K. Jha and S. J. Wang, 
{\it Testing and Reliable Design of CMOS Circuits.}  
(Kluwer, Nowell, MA, 1991).

\item
K.L. Cooke and Z.Grossman, {\it J. Math. Analysis and Applications}
{\bf 86}, 592 (1982);

\item
U. K\"{u}chler and B. Mensch, {\it Stochastics and Stochastics Reports} 
{\bf 40}, 23 (1992).

\item
C. M. Marcus and R. M. Westervelt, {\it Phys. Rev. A} 
{\bf 39}, 347 (1989);

\item
J. A. Hertz, A. Krogh, and R. G. Palmer,
{\it Introduction to the Theory of Neural Computation}  
(Addison--Wesley, Redwood City, CA, 1991).

\item
M. W. Derstine, H. M. Gibbs, F. A. Hopf and D. L. Kaplan.
Phys. Rev. A {\bf 26}, 3720 (1982); 

\item
K. Pyragas.  Phys. Lett. A {\bf 170}, 421 (1992);

\item
T. Ohira,  Phys. Rev. E {\bf  55},  R1255 (1997).

\item
For description of DES, see e.g.,
A. J. Menezes, P. van Oorschot, and S. A. Vanstone, 
{\it Handbook of Applied Cryptography}  (CRC Press, 1996).
A $2^{56}$ bit key scheme was recently revealed
in an open challenge using many computers over the Internet.
(http://www.rsa.com/pressbox/html/980226.html).

\item
The search space 
increases rapidly in particular with the increase of $N$ due to factorials.

\end{enumerate}
\end{itemize}

\end{document}